\documentclass[
aip,
reprint
]{revtex4-2}

\usepackage{graphicx}
\usepackage{amsmath,amssymb}
\usepackage{siunitx}
\usepackage{braket}
\usepackage{dsfont}

\usepackage{hyperref}
\usepackage{xcolor}
\usepackage{cancel} 
\usepackage{floatrow}
\usepackage[caption=false, font = normalsize]{subfig}

\begin{document}
\title{Demonstration and modelling of time-bin entangled photons from a quantum dot in a nanowire}
%
\newcommand{\innsbruckTheory}{Institute for Theoretical Physics, University of Innsbruck, Innsbruck, Austria}
\newcommand{\innsbruckExperiment}{Institute for Experimental Physics, University of Innsbruck, Innsbruck, Austria}
\newcommand{\parityqc}{Parity Quantum Computing GmbH, Innsbruck, Austria}
\newcommand{\ottawa}{National Research Council of Canada, Ottawa, Canada}
\author{Philipp Aumann}
\thanks{These authors contributed equally to this work.}
\affiliation{\innsbruckTheory}
\author{Maximilian Prilmüller}
\thanks{These authors contributed equally to this work.}
\affiliation{\innsbruckExperiment}
\author{Florian Kappe}
\affiliation{\innsbruckExperiment}
\author{Laurin Ostermann}
 \thanks{Author to whom correspondence should be addressed: laurin.ostermann@uibk.ac.at}
\affiliation{\innsbruckTheory}
\author{Dan Dalacu}
\affiliation{\ottawa}
\author{Philip J. Poole}
\affiliation{\ottawa}
\author{Helmut Ritsch}
\affiliation{\innsbruckTheory}
\author{Wolfgang Lechner}
\affiliation{\innsbruckTheory}
\affiliation{\parityqc}
\author{Gregor Weihs}
\affiliation{\innsbruckExperiment}

\date{\today}
\begin{abstract}
Resonant excitation of the biexciton state in an InAsP quantum dot by a phase-coherent pair of picosecond pulses allows preparing time-bin entangled pairs of photons via the biexciton-exciton cascade. We show that this scheme can be implemented for a dot embedded in an InP nanowire. The underlying physical mechanisms can be represented and quantitatively analyzed by an effective three-level open system master equation. Simulation parameters including decay and intensity dependent dephasing rates are extracted from experimental data, which in turn let us predict the resulting entanglement and optimal operating conditions.
\end{abstract}

\maketitle
In a future quantum world long-distance quantum communication will allow users to communicate in perfect privacy and it will connect quantum computers for distributed and blind computation tasks. Quantum repeaters~\cite{Briegel:1998aa} will be necessary in order to establish the required long-distance entanglement and for building even the simplest quantum repeaters we will need reliable, high-rate and high-fidelity sources of entangled photon pairs besides quantum memories and local quantum processing. The emitted photon pairs must propagate with low loss and low decoherence in order to cover as much a distance as possible. While the propagation loss in optical fibers is limited by intrinsic material properties, the decoherence can be minimized by choosing a suitable quantum information encoding~\cite{Tittel:2001aa}. Time-bin entanglement~\cite{Franson:89,Tittel:1998aa} has emerged as the optimal encoding for optical fiber quantum communication, because it is immune to residual fiber birefringence as well as thermal and mechanical fluctuations up to very high frequencies.

\begin{figure}[!t]
\includegraphics[width=1\linewidth]{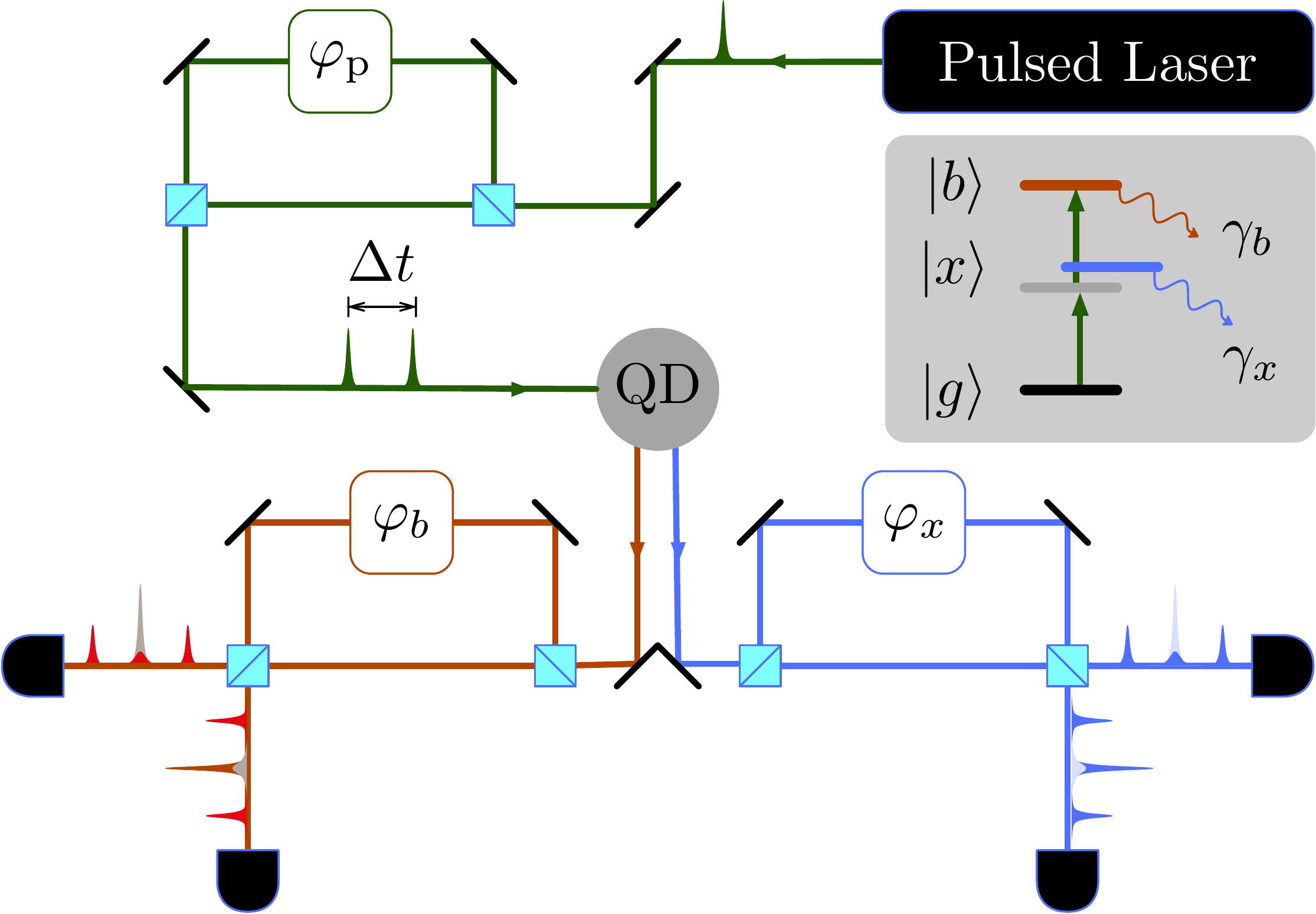}
\caption{\textit{Scheme of the Time-Bin Entanglement Setup}. Three phase-stable interferometers facilitate the generation and projection of time-bin entangled states. The delay of the pump interferometer, $\Delta t$ is chosen to be much longer than the coherence time of the emitted photons to rule out photonic first-order interference. The phases of the three interferometers $\varphi_\mathrm{P}$, $\varphi_b$ and $\varphi_x$ are controlled via phase plates. Each pump pulse excites the system with very low probability in order to ensure that on average maximally one photon pair is created. The interference of these two time bins can be observed when looking at the coincident photon detections between outputs of the different analysis interferometers. Inset: Quantum dot as a three level system (without dark states). Green arrows indicate the direct population of the biexciton state ($\ket{b}$) via a virtual level (gray line). The single photon transition is detuned from the exciton state ($\ket{x}$). Relaxation into the groundstate ($\ket{g}$) results in the successive emission of two photons at different wavelengths.}
\label{fig:Setup}
\end{figure}
So far all sources of time-bin entanglement have been probabilistic, even the ones that used single quantum dots\cite{Jayakumar:2014aa, Versteegh:2015aa}. Most work on quantum dots as entanglement sources has concentrated on maximizing polarization entanglement, for which elaborate growth and tuning techniques have been developed~\cite{Huber:2018ab}. Polarization entanglement can be converted probabilistically to time-bin entanglement~\cite{Versteegh:2015aa} or by using ultra high-speed optical modulators, which, however, are always very lossy and thus do not allow a near-deterministic source. Therefore we consider the direct creation of single time-bin entangled photon pairs from semiconductor quantum dots an important goal. The only known way to achieve this involves using at least three energy levels in the quantum dot, one of which must be metastable~\cite{Simon:2005aa}. While research into deterministic time-bin entanglement from quantum dots is on the way in our laboratory, in this letter, as an intermediate step, we present the realization of probabilistic time-bin entanglement from a quantum dot in an optimized photonic structure.

\begin{figure*}[ht] 
\centering
\sidesubfloat[]{\includegraphics{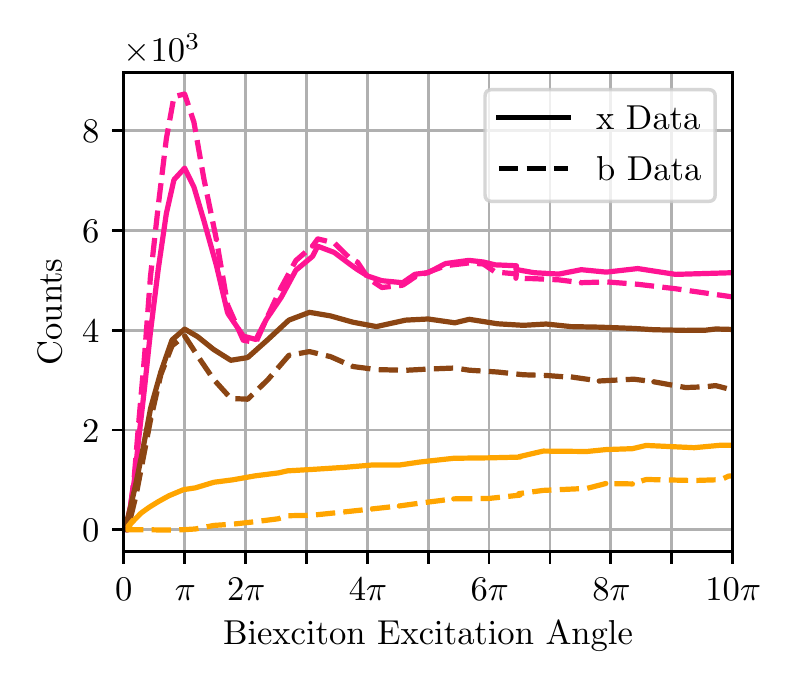}\label{fig:rabi_data}} \hfill
\sidesubfloat[]{\includegraphics{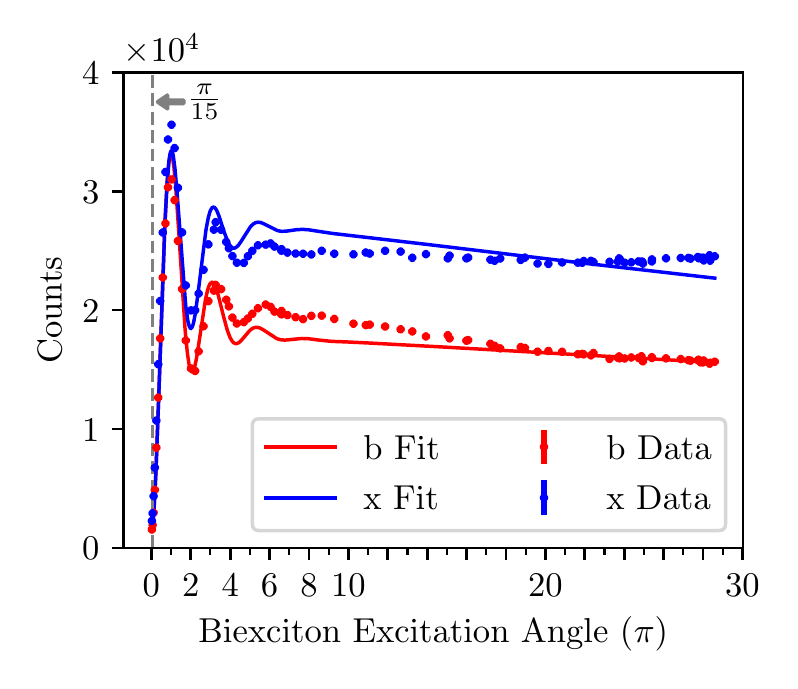}\label{fig:rabi_fit}}
\caption{\textit{Rabi Oscillations of a Quantum Dot Embedded in a Nanowire and Fit of the Emission Probabilities to the Photon Counts}
(a) The solid lines correspond to emission collected at the exciton wavelength, while the dashed lines correspond to biexciton emission, respectively. The horizontal axis represents the average laser power scaled such that the first maximum of the observed oscillations occurs at $\pi$. Pink: \SI{85}{\pico\second} FWHM linearly polarized pump. Brown: \SI{20}{\pico\second} FWHM linearly polarized pump. Orange: \SI{20}{\pico\second} FWHM circular polarized pump, scaled to the biexciton excitation angle of the brown curve. (b) We fitted the emission probabilities predicted by the theoretical model to biexciton and exciton emission counts for a pulse length of $85$~ps FWHM. The resulting parameter values can be found in section~S3 of the supplemental material. The dashed line indicates the position of the $\pi/15$ power that has been used for the time-bin measurement. The measurement error is estimated by the square root of the number of the counts resulting in error bars smaller than the symbols.}
\label{fig:rabi}
\end{figure*}
In the past two decades a lot of work has gone into improving the out-coupling efficiencies of photons from quantum dots~\cite{senellart2017high} e.g. via the implementation of circular Bragg gratings~\cite{wang:2019}, enhancing emission into a collectable mode. Alternatively, realizing quantum dots embedded in tapered nanowires turned out to be a promising platform for coherent photon emission~\cite{Reimer:2016aa, Reimer_NComm, Huber:2014, Versteegh:2014aa}. The tapered part of the nanowire acts as an antenna that matches the impedance of the nanowire waveguide mode to the vacuum and thus achieves efficient out-coupling~\cite{bulgarini_nanowire_2014}.

In the following, we report the generation of time-bin entangled photon pairs in indium arsenide phosphide (InAsP) quantum dots embedded in indium phosphide (InP) nanowires via a resonant two-photon excitation~\cite{koong_multiplexed_2020, basso_basset_entanglement_2019} (see figure~\ref{fig:Setup}). Furthermore, we present an extension of our theoretical model from previous work~\cite{Huber:2016} that includes the density matrix of the time-bin entangled photons, which allows suggesting optimal  parameter values.

\textit{Experiment} -- The core of our setup is constituted by a quantum dot embedded in a nanowire. Our samples were manufactured utilizing a selective-area vapor-liquid-solid epitaxy which produced InAsP quantum dots embedded in defect-free wurtzite InP nanowires~\cite{Dalacu:2012}. A single electron-hole pair trapped in the quantum dot is referred to as an exciton ($\ket{x}$), while the confinement of two pairs is called a biexciton ($\ket{b}$). A recombination of a single pair leads to the emission of a photon at a characteristic wavelength, as depicted in the inset of figure~\ref{fig:Setup}. The biexciton-exciton photon cascade is used in order to operate the quantum dot as a source of correlated photon pairs. The emission spectrum of our quantum dot can be found in figure~S1 in the supplementary material. 

\begin{figure*}[t]
\centering
\includegraphics{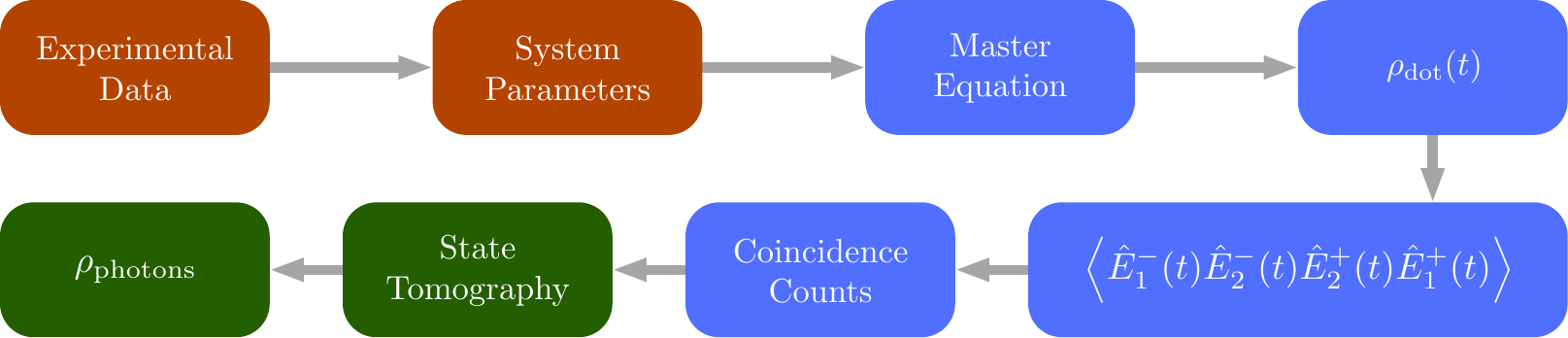}
\caption{\textit{Procedure for Simulating the Photonic Density Matrix from the Quantum Dot Dynamics}. After setting our model parameters to fit the experimental Rabi data, we simulate the dynamics of the quantum dot's density matrix, obtaining the photon coincidence counts via calculating the two-photon detection probabilities and thus reconstruct the photonic density matrix by means of state tomography. More details on the procedure to estimate the density matrix of the photons $\rho_{\text{photons}}$ from our theoretical model are given in section~S2 of the supplementary material.}
\label{fig:Concept}
\end{figure*}
The basic principle of the time-bin encoding scheme relies on the ability to create a coherent superposition of two well-defined excitation processes. Its simplest realisation relies on addressing the emitter with two pump pulses of very low excitation probability each, and postselecting on biexciton - exciton photon pair detection events. The two excitation pulses are created in an unbalanced Mach-Zehnder interferometer and denoted by \textit{e} (early) and \textit{l} (late). The phase between the two pulses $\Delta_\phi$ can be modified via a phase plate and determines the phase of the entangled state. Denoting biexciton and exciton photons by $b$ and $x$ respectively, the created state can be written as:
\begin{align}
\ket{\Phi} &=   \frac{1}{\sqrt{2}}  \Bigl(  \ket{e}_{b} \ket{e}_x + e^{i \Delta_{\phi}} \ket{l}_{b} \ket{l}_x \Bigr) \nonumber \\
			  &=:  \frac{1}{\sqrt{2}}  \left( \ket{ee}  + e^{i \Delta_{\phi}} \ket{ll} \right).
\end{align}

Using another two unbalanced Mach-Zehnder interferometers that are phase stable with respect to the pump interferometer we carry out projective measurements on the created entangled state. In order to perform quantum state tomography, we analyze the result of $16$ different pairs of phase settings and use a maximum likelihood approach.~\cite{James:2001aa,Takesue:2009aa} For collecting the $16$ different projections necessary for the quantum state tomography we employ four different phase settings in the analysis interferometers each and detect photons at each of the four output ports. We collect time tags of the detected photons for \SI{3600}{s} per phase setting and identify coincident photon pairs by imposing a coincidence window of \SI{400}{ps}. The integration time was chosen such that it would yield sufficient statistics for the maximum likelihood reconstruction method~\cite{Schwemmer:2015aa}.

For the generation of biexciton-exciton photon pairs, we employ resonant pulsed two-photon excitation from $\ket{g}$ to $\ket{b}$ (see inset in figure~\ref{fig:Setup}). In order to calibrate and characterize the system, we observe Rabi oscillations by measuring the photon counts as a function of the average laser power as shown in figure~\ref{fig:rabi_data}. We see that it is critical to identify an appropriate polarization as well as a sensible pulse duration. Choosing a circular pump polarization violates optical selection rules and leads to incoherent excitations rather than to a two-photon coherent coupling of the ground and biexciton state. By comparing the oscillations resulting from a linearly polarized pump and pulse lengths of $25$~ps and $85$~ps, we find a significantly stronger coherence for the longer pulse. The similar slopes at low excitation power of the biexciton and exciton emission probabilities for a linearly polarized pump indicate the superior pair production efficiency of this excitation scheme.

For the creation of time-bin entangled photons we thus use the optimized pulse duration of $85$~ps~\cite{Huber:2016} resulting in a substantial increase of the excitation coherence and we determine the energy of a $\pi/15$-pulse to be adequate, yielding an excitation probability of about $7.5$~$\%$ per pulse which reduces the probability of emitting at both time bins to below $0.6$~$\%$. Our theoretical model (see below) underpins the feasibility of the chosen parameters and provides the basis for even further improvements in future work.

\begin{figure*}[t]
\centering
\includegraphics{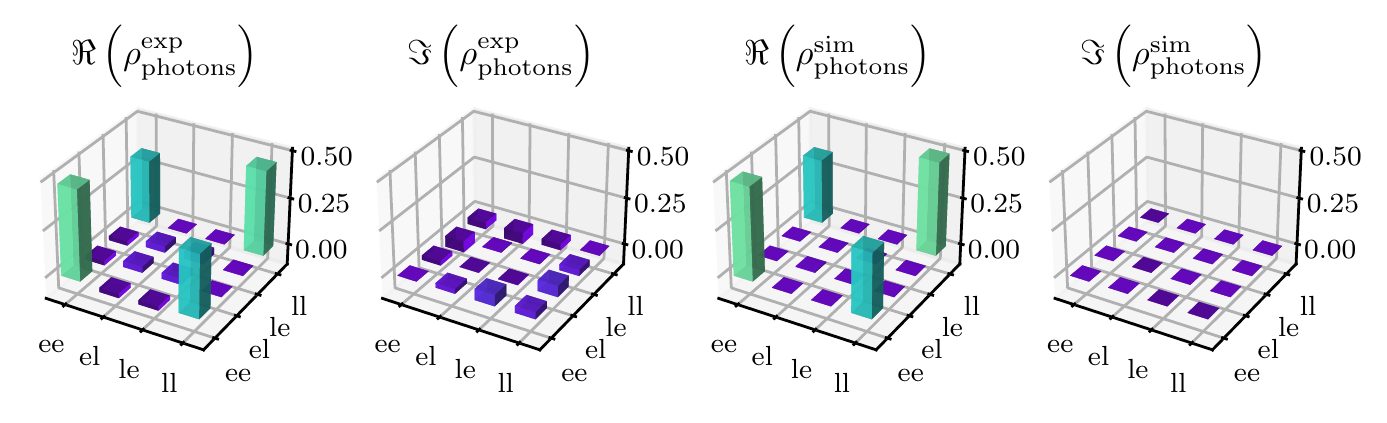}
\caption{\textit{Reconstructed Density Matrix of the Emitted Photons}. Left: Real and imaginary part of the reconstructed density matrix from the experiment $\rho_{\mathrm{photons}}^{\mathrm{exp}}$.
Right: Real and imaginary part of the simulated density matrix $\rho_{\mathrm{photons}}^{\mathrm{sim}}$. The agreement between the experimentally obtained and the simulated density matrix is calculated by means of equation~\ref{eq:densityfidelity} yielding a value of around $F_{\rho} \approx 0.96$.}
\label{fig:density_matrices}
\end{figure*}
\textit{Theoretical Model} -- We propose a quantum optical model in order to identify a suitable set of parameter values to enhance the quality of entanglement of the created photon pairs. This allows us to estimate the reconstructed density matrix as shown in figure~\ref{fig:density_matrices}. Extending our earlier work~\cite{Huber:2016}, where we used a model for the internal dynamics of the quantum dot, we include a procedure for obtaining the photons' density matrix from said quantum dot dynamics. Our strategy is outlined in figure~\ref{fig:Concept}. The conceptual procedure introduced here is not restricted to this particular experimental setup and thus can be seen as a more fundamental framework for a general setting of time-bin entangled photons from a quantum dot. 

The pulsed laser driving field couples the ground state to the biexciton via the ground state to exciton and exciton to biexciton transition dipoles. The Hamiltonian in the effective interaction picture reads (for a derivation see section~S4 of the supplementary material)
\begin{equation}\label{eq:H}
\begin{aligned}
H =	& ( \Delta_x - \Delta_b ) \ket{x} \bra{x} -2 \Delta_b \ket{b} \bra{b} \\
		& + \Omega (t)  \Bigl( \ket{g}\bra{x} + \ket{x}\bra{b} + \mathrm{h.c.} \Bigr).
\end{aligned}
\end{equation}

Here, $\Delta_x$ denotes the detuning from the exciton level to the laser frequency, while $\Delta_b$ is the detuning from the biexciton level to the two-photon transition, $\Omega (t)$ is the Rabi frequency featuring a Gaussian time profile,
\begin{equation}\label{eq:RabiFrequency}
\Omega (t)=\Omega_0 \exp \left( -\frac{4\ln (2)(t-t_0)^2}{\tau^2} \right),
\end{equation}
with amplitude $\Omega_0$, pulse duration (FWHM) $\tau$ and time offset $t_0$.

To simulate the dynamics we solve the master equation in Lindblad form numerically, i.e.\
\begin{equation}\label{eq:MELindblad}
\dot \rho = i \left[ \rho, H \right] + \frac{1}{2} \sum_{j=1}^6 \left( 2 R_j \rho R_j^\dagger - R_j^\dagger R_j \rho - \rho R_j^\dagger R_j \right).
\end{equation}
where $\rho =\rho_{\mathrm{dot}}(t)$ is the quantum dot density matrix. We consider six dissipative channels associated with six different Lindblad operators $R_j$, where
\begin{align}
R_1 =& \sqrt{\gamma_x} \ \ket{g}\bra{x}, \\
R_2 =& \sqrt{\gamma_b} \ \ket{x}\bra{b},
\end{align}
describes the radiative decay of the biexciton and exciton levels with rates $\gamma_b$ and $\gamma_x$, respectively, while
\begin{align}
R_3 =& \sqrt{\gamma^{\text{const}}_{xg} + \gamma_{xg}} \ (\ket{x}\bra{x} - \ket{g}\bra{g}) \\
R_4=& \sqrt{\gamma^{\text{const}}_{bx} + \gamma_{bx}} \ (\ket{b}\bra{b} - \ket{x}\bra{x})
\end{align}
introduce dephasing. The rates ${\gamma_{bx} = \gamma^{I_0}_{bx}\Bigl( \tfrac{\Omega(t)}{\Omega_S} \Bigr)^{n}}$ and ${\gamma_{xg} = \gamma^{I_0}_{xg}\Bigl( \tfrac{\Omega(t)}{\Omega_S} \Bigr)^{n}}$ are comprised of their amplitudes $\gamma^{I_0}_{bx}$ and $\gamma^{I_0}_{xg}$ as well as the scaled time-dependent Rabi frequency to the $n$-th power. Throughout this work we consider $n=2$. This laser intensity dependent dephasing can be explained by phonons coupling to the quantum dot~\cite{ramsay_damping_2010}.  The scaling factor ${\Omega_S = 1 }$~THz accounts for the correct numerical values and leads to a unitless expression for the Rabi frequency. A minor role is attributed to the rates of constant dephasing $\gamma^{const}_{xg}$ and $\gamma^{const}_{bx}$ by the fit in figure~\ref{fig:rabi_fit}. 

In order to account for the decrease of photon counts for higher laser power as depicted in figure~\ref{fig:rabi_fit}, we introduce dark states modelling a laser power dependent loss mechanism, as states outside the three-level approximation become more prominent for higher laser powers. Moreover, this additional dark state loss counteracts the increased exciton population via a single photon transition that appears at higher laser intensities based on the broadening of the spectral linewidth due to the laser dependent dephasing. For bookkeeping purposes, we introduce two dark states $\ket{d_x}$ and $\ket{d_b}$, which are populated by laser dependent exciton and biexciton decay, whereas in general one dark state would suffice to constitute the same effect. The corresponding Lindblad operators are given by
\begin{align}
R_5 =& \sqrt{\gamma_{xd}} \ \ket{d_x}\bra{x}, \\
R_6 =& \sqrt{\gamma_{bd}} \ \ket{d_b}\bra{b},
\end{align}
with laser intensity dependent decay rates ${\gamma_{xd} = \gamma^{I_0}_{xd}\Bigl(\tfrac{\Omega(t)}{\Omega_S} \Bigr)^{n}}$ and ${\gamma_{bd} = \gamma^{I_0}_{bd}\Bigl(\tfrac{\Omega(t)}{\Omega_S} \Bigr)^{n}}$, decay amplitudes $\gamma^{I_0}_{xd}$ and $\gamma^{I_0}_{bd}$, as well as the same power $n$ as the dephasing mechanism.

Exemplary dynamics of the quantum dot when driven by a laser pulse are depicted in figure~S2 and numerical values for the system parameters can be found in table~S2 and S3 of the supplementary material.

\begin{figure*}[ht]
\centering
\sidesubfloat[]{\includegraphics{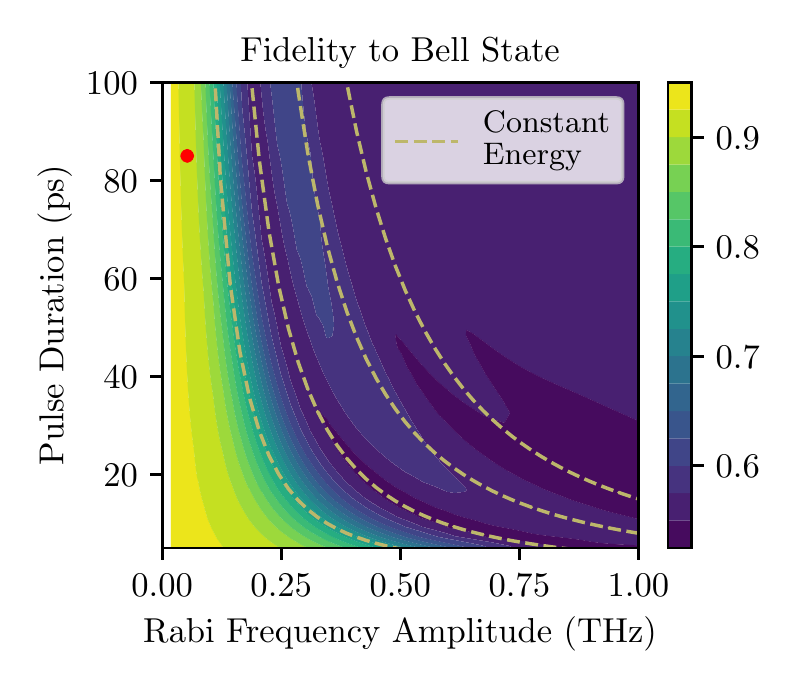}\label{fig:fidelity_contour}} \hfill
\sidesubfloat[]{\includegraphics{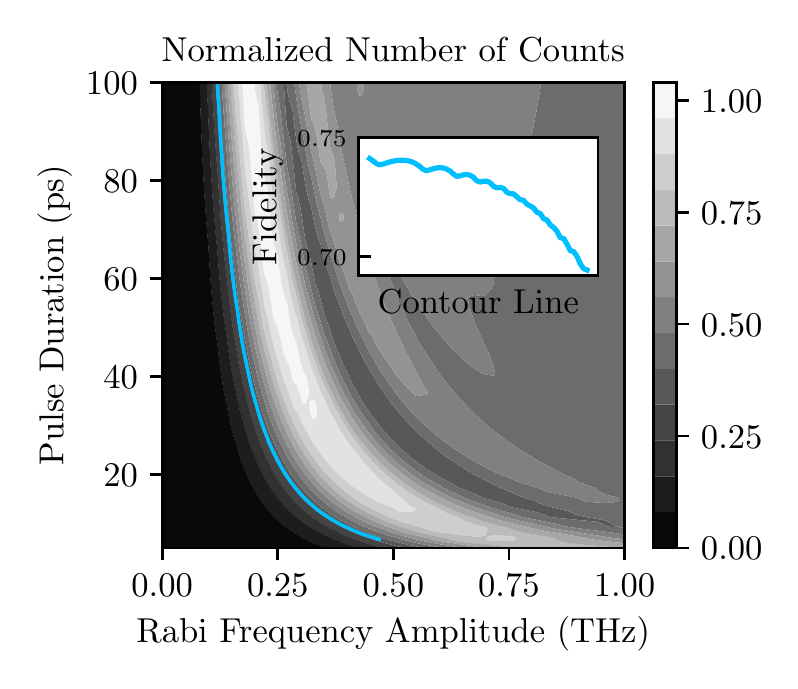}\label{fig:fidelity_counts}}
\caption{\textit{Fidelity with the Bell State and Normalized Number of Photon Counts}. (a) Fidelity $F_{\ket{\Phi^+}}$ (see equation~\ref{eq:FidelityPureState}) between the theoretically predicted density matrix $\rho_{\mathrm{photons}}$ and $\ket{\Phi^+}$ for various values of Rabi frequency amplitude $\Omega_0$ and the pulse duration $\tau$. The red dot marks the parameters chosen for figure~\ref{fig:density_matrices}, i.e. an excitation angle of $\pi/15$ ($\Omega_0 \approx 0.05$) and a pulse duration of $85$~ps FWHM for the measurement of time-bin entangled photons. The dashed lines indicate areas of constant pulse energy proportional to $\Omega_0^2 \tau$. (b) Normalized number of total counts predicted by the simulated projective measurements. The blue line in the main plot follows a constant count number of $0.32$. Additionally, the inset depicts the fidelity along this line when going from small to large $\Omega_0$.
}
\end{figure*}
In order to obtain the density matrix of the emitted photons from the quantum dot dynamics, we mimic the measurement of the photon coincidence counts in the experiment: first, we calibrate our model by fitting the emission probabilities 
\begin{align}\label{eq:EmissionProbabilitiesMainText}
P_{i} = \gamma_{i} \int \bra{i} \rho_{\mathrm{dot}}(t) \ket{i} \mathrm{d} t  
\end{align}
for $i \in \lbrace x, b \rbrace$ to the biexciton and exciton Rabi data (see figure~\ref{fig:rabi_fit}). This comprehensive and numerical demanding fitting loop is outlined in section~S2 of the supplemental material. Then, the density matrix of the quantum dot as a function of time $\rho_{\mathrm{dot}}(t)$ is the result of numerically solving the master equation. Relating the density matrix of the photons $\rho_{\mathrm{photons}}$ to $\rho_{\mathrm{dot}}(t)$ is achieved by calculating the resulting photon coincidence counts, where we derive analytic expressions for the detection probabilities of all $16$ projective measurements (see~S2 in the supplement) and subsequently use this estimate as an input for the conventional state tomography. This procedure is depicted schematically in figure~\ref{fig:Concept}. The density matrix resulting from this approach is shown in figure~\ref{fig:density_matrices} and compared to the density matrix obtained in the experiment. In order to quantitatively compare the experimental results to our simulation we employ to following definition of the fidelity for two mixed states~\cite{nielsen_quantum_2010}:
\begin{align}\label{eq:densityfidelity}
    F_{\rho} = \mathrm{tr} \ \sqrt{ \sqrt{\rho_{\mathrm{photons}}^{\mathrm{exp}}} \ \rho_{\mathrm{photons}}^{\mathrm{sim}} \ \sqrt{\rho_{\mathrm{photons}}^{\mathrm{exp}}} } \approx 0.96.
\end{align}

The density matrix from theory and experiment have the same structural appeareance, as they show similar values at the prominent matrix elements. While the remaining entries of the simulated density matrix appear to be rather flat, we observe additional small fluctuations of these entries for the density matrix from the experiment.

\textit{Entanglement Quality of the Photons} -- Ultimately, our goal is to achieve two-photon emission in a perfect Bell state,
\begin{align}
    \ket{\Phi^+}= \frac{1}{\sqrt{2}} \Bigl( \ket{ee} + \ket{ll} \Bigr).
\end{align}

Therefore, we identify suitable values for the laser intensity $I \propto \Omega^2$ and its pulse duration $\tau$ in our simulation, which can assist in maximizing the fidelity,
\begin{align}\label{eq:FidelityPureState}
F_{\ket{\Phi^+}} = \sqrt{| \bra{\Phi^+} \rho_{\mathrm{photons}} \ket{\Phi^+} |},
\end{align}
to a Bell state in the experiment. Figure~\ref{fig:fidelity_contour} shows a scan of the fidelity $F_{\ket{\Phi^+}}$ over the corresponding parameter space spanned by $\Omega_0$ and $\tau$. Here, we study the influence of the parameter $\Omega_0$ instead of the intensity $I$ as this parameter is more natural to the theoretical model. Once the model is calibrated to the experimental data, $\Omega_0$ can be converted to the average laser power.

Similar to the Rabi oscillations in figure~\ref{fig:rabi}, we observe an oscillatory pattern, which becomes less and less pronounced towards regions of higher energy (upper right corner). This can mostly be attributed to the intensity-dependent dephasing. For lower energies (lower left corner) the pattern roughly follows areas of constant energy, indicated by the yellow dashed lines. The red dot indicates the values chosen in the measurements that yield the reconstructed time-bin encoded photonic density matrix in figure~\ref{fig:density_matrices}. We show simulated density matrices for the same pulse length but different average laser power in figure~S4 of the supplemental material, where we observe an increase of the diagonal entries of the density matrix towards regions of lower fidelity which means that the photonic state is becoming more classical in low fidelity regions for this pulse length. Reaching the regime of maximal fidelity has to be deferred to a future experimental setup, where our theoretical model can prove even more useful in fine-tuning the experimental parameters.

For a source of entangled photons it is desirable to not only achieve a high fidelity, but also to yield sufficient output. Figure~\ref{fig:fidelity_counts} depicts the normalized number of total expected counts of all simulated projective measurements (see supplementary material~S2). Again, we observe an oscillatory behaviour where we find some degree of anti-correlation between the pattern of the counts and the fidelity, i.e.\ dark areas with less output correspond to a relatively high fidelity, whereas bright areas are connected to a smaller fidelity. Yet, these two patterns are not perfectly anti-correlated, as we find slightly varying fidelity for contours of constant counts. For some applications, a minimum amount of photons is required. Consequently, one might be interested in the optimal fidelity for a given photon count rate. For instance, we observe the fidelity along a contour of constant counts in the inset of figure~\ref{fig:fidelity_contour}. For this particular contour, we find the highest fidelity for long pulses with a relatively low intensity. In cases where the rate of output photons is not an issue, our study suggests that the optimal parameter regime is that of low pulse energy (lower left corner).

\textit{Conclusions} -- In this work we have shown the coherent coupling of ground to biexciton state of a InAsP quantum dot embedded in an InP nanowire via an optimised two-photon resonant excitation scheme. We have used this method to generate time-bin entangled photons, yielding a fidelity of $F_{\ket{\Phi^+}} \approx 0.90$ (see equation~\ref{eq:FidelityPureState}) with respect to the maximally entangled $\ket{\Phi^{+}}$ Bell state. 

Additionally, we have presented a quantum optical model for simulating the dynamics of the quantum dot. By making use of the experimental reconstruction method, we have introduced a scheme for predicting the density matrix of the emitted photons based on the simulation of the dynamics of the quantum dot. The results of the model have been compared to the outcome of the experiment. With this, we are able to identify optimal parameter regimes in order to further increase the fidelity of the photons' density matrix to a Bell state and to provide a more general toolbox for the study of time-bin entangled photons from a quantum dot. 

\section*{Acknowledgements}
We want to thank Doris Reiter and her group for fruitful discussions. P.A.\ and W.L.\ are supported by the Austrian Science Fund (FWF) through a START grant under Project No. Y1067-N27 and the SFB BeyondC Project No. F7108, the Hauser-Raspe foundation and the European Union's Horizon 2020 research and innovation program under grant agreement No. 817482. M.P.\, F.K.\, and G.W.\ acknowledge partial support by the Austrian Science Fund (FWF) through projects W1259 (DK-ALM), I4380 (AEQuDot), and F7114 (SFB BeyondC). This material is based upon work supported by the Defense Advanced Research Projects Agency (DARPA) under Contract No. HR001120C0068. Any opinions, findings and conclusions or recommendations expressed in this material are those of the author(s) and do not necessarily reflect the views of DARPA.

\section*{Supplementary Material}
The supplementary material shows the quantum dot's emission spectrum and details on the experimental methods. It features exemplary dynamics of the quantum dot upon excitation by a laser pulse and provides an in-depth mathematical assessment of the reconstruction of the photons' density matrix from the quamtum dot's density matrix. Furthermore, it contains a summary of the chosen values for the simulation parameters including the fit of the decay rates and a derivation of the Hamiltonian in equation~\ref{eq:H}.

\section*{Data availability}
The data that support the findings of this study are available from the corresponding author upon reasonable request.

\section*{Conflict of Interest}
The authors have no conflicts to disclose.

\section*{References}
\bibliographystyle{unsrt}

\end{document}


\title{Supplementary Material: Demonstration and modelling of time-bin entangled photons from a quantum dot in a nanowire}

\newcommand{\innsbruckTheory}{Institute for Theoretical Physics, University of Innsbruck, Innsbruck, Austria}
\newcommand{\innsbruckExperiment}{Institute for Experimental Physics, University of Innsbruck, Innsbruck, Austria}
\newcommand{\parityqc}{Parity Quantum Computing GmbH, Innsbruck, Austria}
\newcommand{\ottawa}{National Research Council of Canada, Ottawa, Canada}

\author{Philipp Aumann}
\thanks{These authors contributed equally to this work.}
\affiliation{\innsbruckTheory}
\author{Maximilian Prilmüller}
\thanks{These authors contributed equally to this work.}
\affiliation{\innsbruckExperiment}
\author{Florian Kappe}
\affiliation{\innsbruckExperiment}
\author{Laurin Ostermann}
 \thanks{Author to whom correspondence should be \\ addressed: laurin.ostermann@uibk.ac.at}
\affiliation{\innsbruckTheory}
\author{Dan Dalacu}
\affiliation{\ottawa}
\author{Philip J. Poole}
\affiliation{\ottawa}
\author{Helmut Ritsch}
\affiliation{\innsbruckTheory}
\author{Wolfgang Lechner}
\affiliation{\innsbruckTheory}
\affiliation{\parityqc}
\author{Gregor Weihs}
\affiliation{\innsbruckExperiment}


\maketitle
\onecolumngrid

In this supplementary material we provide more details about the experimental methods in section~\ref{sec:Methods}. In section~\ref{sec:RhoPhotonsEstimation}, we describe the theoretical procedure to estimate the density matrix of the photons from the simulation of the quantum dot. Additionally, we give an overview of the numerical values for the system parameters in section~\ref{sec:SystemParameters} and close with a derivation of the Hamiltonian in the final section~\ref{sec:Hamiltonian}.

\section{Experimental Methods}\label{sec:Methods}
%
%
When using pulsed excitation schemes the temporal envelope of the pump pulses has to be optimized: Short pulses would give an ideal timing of the excitation process, but the high peak intensity of the driving field has detrimental effects on the quantum dot system~\cite{Huber:2016}. Moreover, the increased pump bandwidth would reduce the selectivity of the driving field and could give rise to a direct excitation of excitonic states. The phenomenological model developed by Huber et al.~\cite{Huber:2016} has inspired the use of the particular excitation pulse shapes in this work.

\begin{figure}[b]
\centering
\includegraphics{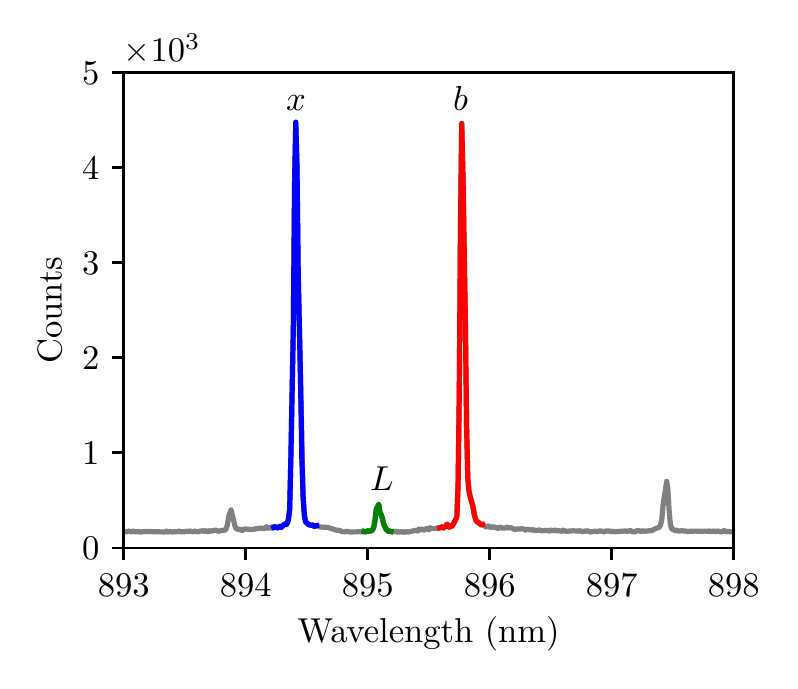}
\caption{\textit{Photon Counts Recorded on a Spectrometer}. The lines are assigned to the exciton ($x$) and biexciton ($b$) emission wavelengths. In order to fulfill the energy conservation for a resonant two-photon excitation, the pump's wavelength ($L$) is tuned to be centered between the two sharp emission lines. Reflected laser light is suppressed by two narrow-band notch filters in front of the spectrometer but can still be seen in the spectrum.}
\label{fig:spectrum}
\end{figure}

Furthermore, the two-photon resonance imposes optical selection rules on the driving field, such that it needs to feature linear polarizations exclusively~\cite{Flissikowski:2004aa}. The requirement of actively choosing the pump polarization did not arise in our earlier works using planar cavities, as the laterally coupled waveguide would effectively act as a polarizer~\cite{Jayakumar:2013, Huber:2016aa}. Using a collinear pump/probe setup however, we are able to demonstrate the adverse effects of inadequately polarized pump light: Rather than a direct two-photon excitation of the biexciton we observe consecutive excitation of the exciton state (orange line in figure~2a of the main text).

A simultaneous optimization of bandwidth, pulse duration and polarization of the two-photon resonant driving field enables the realization of time-bin entangled pairs of photons emitted by a quantum dot~\cite{Jayakumar:2014aa} embedded in a nanowire.

The nanowire sample is kept inside a liquid helium flow cryostat set to \SI{5}{\kelvin}, and optically accessed along the optical axis of the nanowire by a single aspheric lens ($\mathrm{NA}=0.68$). Taking advantage of the single-mode emitter nature of the nanowire waveguides, a highly efficient colinear pump/probe setup can be employed for these measurements. A combination of half- and quarter-wave plates ensures the alignment of the quantum dot polarization with the experimental setup, which is crucial for addressing the two-photon resonance efficiently~\cite{Chen:2002aa}.

The pump light is prepared in a Gaussian mode of a well defined polarization, and inserted into the optical path by means of a 90/10 beam splitter. A pair of two narrow-band notch filters is put into the optical collection path in order to suppress the residual backscattered pump light.

The quantum dot emission is spectrally separated by a grating ($1500$ grooves per millimeter) and coupled into mode-matched single mode fibers. Superconducting nanowire single photon detectors (SNSPDs) detect the collected and filtered photons. A pellicle beam splitter can be inserted into the optical path in order to divert parts of the nanowire emission towards a spectrometer. This is used for the Rabi oscillation analysis performed by integrating over the emission intensities of the individual quantum dot lines for varying pump powers.

For a pulsed excitation the coherent pump pulses are generated by an actively mode locked titanium sapphire pulsed laser (\SI{81}{\mega\hertz} repetition rate), capable of generating pulse durations up to \SI{50}{\pico\second}. In order to reach even longer pulse durations up to \SI{100}{\pico\second}, a pulse stretcher can be inserted into the pump path. Upon optimizing the driving field, a spectral notch filter strongly suppresses the resulting weak pump residue as it is spectrally distinct from the quantum dot emission lines of interest.

\begin{figure}[t]
\centering
\includegraphics{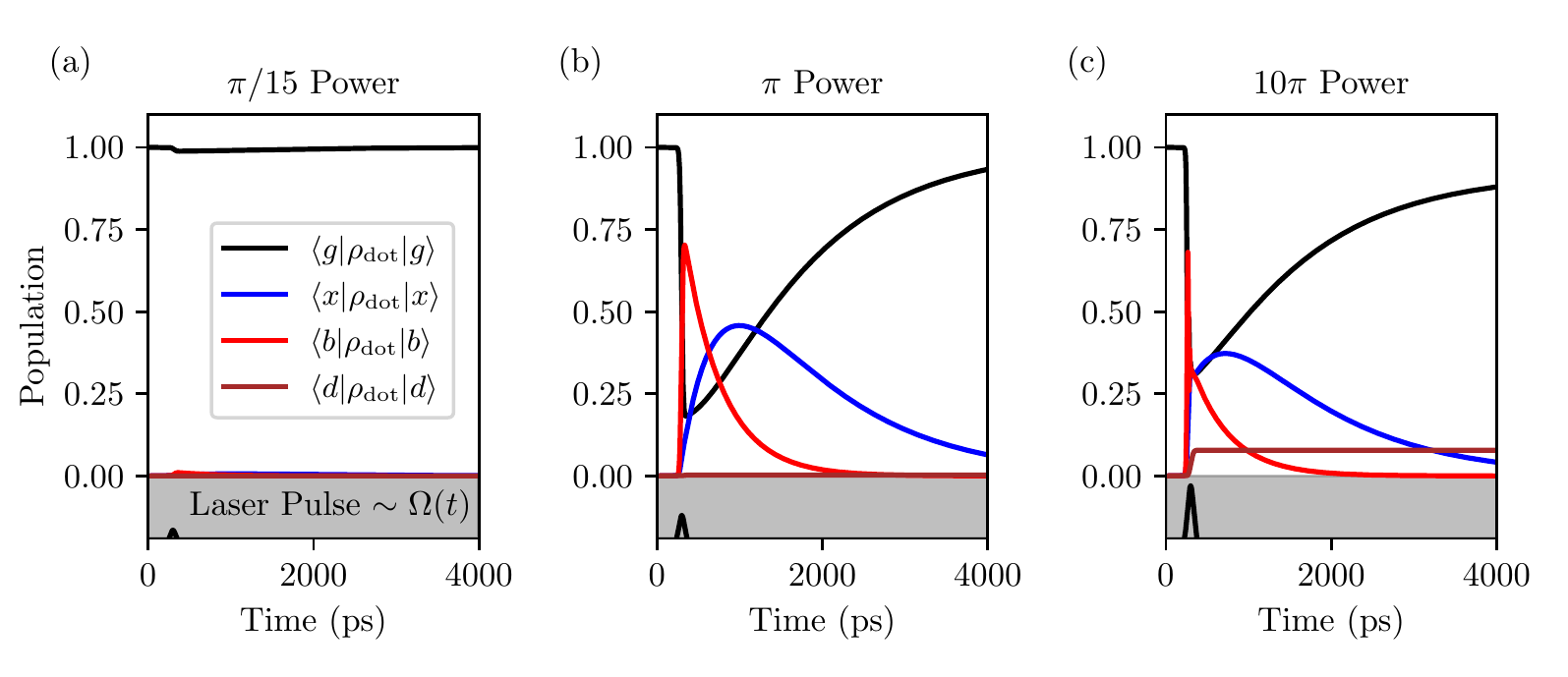}
\caption{\textit{Illustration of the Quantum Dot's Time Evolution}. The population of the two dark states $\ket{d_x}$ and $\ket{d_b}$ is summarized within $\bra{d} \rho_{\textrm{dot}}\ket{d}$. The dynamics is shown for a factor of $\frac{1}{15}$ (a), $1$ (b) and $10$ (c) of to the biexciton excitation angle $\pi$ from figure~2b of the main text. The laser pulse is depicted at the bottom of the figure featuring a pulse length of $85~$ps. The amplitude of the laser pulse has been adjusted for illustrational purposes and is not drawn to scale.}
\label{fig:Dynamics} 
\end{figure}

\section{Photon Density Matrix Estimation}\label{sec:RhoPhotonsEstimation}
An overview of the procedure to deduce the density matrix of the photons $\rho_{\mathrm{photons}}$ from the simulation of the quantum dot system is given in this section. We start by extracting the system parameters from the experiment. This includes calculating the emission probability of an exciton and a biexciton photon given by (as stated in equation~11 in the main text)
\begin{align} 
P_{i} = \gamma_{i} \int \bra{i} \rho_{\mathrm{dot}}(t) \ket{i} dt
\label{eq:EmissionProbabilities}
\end{align}
with  $i \in \{ x,b \}$ and the integral taken over a time span that captures all relevant dynamics. In order to fit the parameters, these probabilities are scaled by biexciton and exciton counts scaling factor $k_{c_{\mathrm{scale}_b}}$ and $k_{c_{\mathrm{scale}_x}}$, provided with offsets $k_{c_{\mathrm{off}_b}}$ and $k_{c_{\mathrm{off}_x}}$ and subsequently compared to the detection counts of biexciton and exciton photons from the experiment as a function of average laser power (see figure~2b of the main text). The fitting loop starts by adjusting the average laser power to a fixed value $p$. This value translates to the amplitude of the Rabi frequency amplitude $\Omega_0$ with a scaling constant $k_{P_\mathrm{scale}}$, power offset $k_{P_\mathrm{off}}$ and pulse length $\tau$ by $\Omega_0 = \sqrt{k_{P_\mathrm{scale}} \frac{|p + k_{P_\mathrm{off}}|}{\tau}}$. Subsequently the density matrix of the quantum dot $\rho_{\mathrm{dot}}(t)$ is calculated by solving the master equation numerically (see equation~4 in the main text). Now, the probabilities of exciton and biexciton emission are estimated by using equation~\ref{eq:EmissionProbabilities} and the resulting value is compared to the measured photon counts. This process is performed for multiple values of the average laser power shown in figure~2b in the main text (the biexciton excitation angle corresponds to the average laser power). In order to estimate the system parameters from the experimental data, this comprehensive fitting loop is repeated until the fitting procedure finishes. The fit is calculated with the support of the Python module \texttt{blackbox}~\cite{knysh_blackbox_2016}.

\begin{table*}[t]
\caption{\linespread{1.1}\selectfont A summary of the states used for simulating the projective measurements and the corresponding probability distributions $\eta$ which are used to estimate the number of coincidence counts. This table is taken from reference~\cite{aumann_tomography_2017}.}
\linespread{1.4}\selectfont
\begin{tabular}{ m{0.03\textwidth}  m{0.05\textwidth}  m{0.05\textwidth}  m{0.3\textwidth} m{0.42\textwidth} }
\hline
$\nu$ & Biex-\newline citon & Ex-\newline citon & Composite State & $\eta_{\nu} \ \text{(proportional)}$  \\ \hline \hline 
1 & $\ket{+}$ & $\ket{+}$ & $\tfrac{1}{2} ( \ket{ee} + \ket{el} + \ket{le} + \ket{ll})$ & $\tfrac{1}{2}(\rho_{bb}(1+ \rho_{xx} ) + 2 \rho_{bx} \rho_{xb} + \rho_{bg} \rho_{gb} + \rho_{xx} \rho_{bb} )$   \\
2 & $\ket{+}$ & $\ket{R}$ & $\tfrac{1}{2} ( \ket{ee} - i \ket{el} + \ket{le} - i \ket{ll})$ & $\tfrac{1}{2}(\rho_{bb}(1+ \rho_{xx} ) + 2 \rho_{bx} \rho_{xb} + \rho_{xx} \rho_{bb} )$  \\
3 & $\ket{+}$ & $\ket{e}$ & $\tfrac{1}{\sqrt{2}} ( \ket{ee} + \ket{le})  $ & $\tfrac{1}{2}(\rho_{bb}(1+\rho_{xx}) + 2 \rho_{bx} \rho_{xb} + \rho_{xx} \rho_{bb})$  \\
4 & $\ket{+}$ & $\ket{l}$ & $\tfrac{1}{\sqrt{2}} ( \ket{el} + \ket{ll})  $ & $\tfrac{1}{2}(\rho_{bb}(1+\rho_{xx}) + 2 \rho_{bx} \rho_{xb} + \rho_{xx} \rho_{bb}) $  \\
5 & $\ket{R}$ & $\ket{+}$ & $\tfrac{1}{2} ( \ket{ee} + \ket{el} -i  \ket{le} - i \ket{ll})$ & $\tfrac{1}{2}(\rho_{bb}(1+ \rho_{xx} )+ \rho_{xx} \rho_{bb} )$  \\
6 & $\ket{R}$ & $\ket{R}$ & $\tfrac{1}{2} ( \ket{ee} - i \ket{el} -i  \ket{le} -  \ket{ll})$  & $\tfrac{1}{2}(\rho_{bb}(1+ \rho_{xx} ) - \rho_{bg} \rho_{gb} + \rho_{xx} \rho_{bb} )$ \\
7 & $\ket{R}$ & $\ket{e}$ & $\tfrac{1}{\sqrt{2}} ( \ket{ee} - i \ket{le})  $ & $ \tfrac{1}{2}(\rho_{bb}(1+\rho_{xx})  + \rho_{xx} \rho_{bb})$ \\
8 & $\ket{R}$ & $\ket{l}$ & $\tfrac{1}{\sqrt{2}} ( \ket{el} -i \ket{ll} ) $ & $ \tfrac{1}{2}(\rho_{bb}(1+\rho_{xx}) + \rho_{xx} \rho_{bb}) $ \\
9 & $\ket{e}$ & $\ket{+}$ & $\tfrac{1}{\sqrt{2}} ( \ket{ee} + \ket{el}) $ & $\tfrac{1}{2}(\rho_{bb}(1+\rho_{xx}) +  \rho_{xx} \rho_{bb})$ \\
10 & $\ket{e}$ & $\ket{R}$ & $\tfrac{1}{\sqrt{2}} ( \ket{ee} -i \ket{el}) $ & $\tfrac{1}{2}(\rho_{bb}(1+\rho_{xx}) +  \rho_{xx} \rho_{bb})$ \\
11 & $\ket{e}$ & $\ket{e}$ & $ \ket{ee} $ & $\rho_{bb}$ \\
12 & $\ket{e}$ & $\ket{l}$ & $ \ket{el} $ & $2 \rho_{xx} \rho_{bb}$ \\
13 & $\ket{l}$ & $\ket{+}$ & $\tfrac{1}{\sqrt{2}} ( \ket{le} + \ket{ll})$ & $\tfrac{1}{2}(\rho_{bb}(1+\rho_{xx}) +  \rho_{xx} \rho_{bb})$  \\
14 & $\ket{l}$ & $\ket{R}$ & $\tfrac{1}{\sqrt{2}} ( \ket{le} - i \ket{ll})$ & $\tfrac{1}{2}(\rho_{bb}(1+\rho_{xx}) +  \rho_{xx} \rho_{bb})$\\
15 & $\ket{l}$ & $\ket{e}$ & $ \ket{le} $ & $2 \rho_{xx} \rho_{bb}$\\
16 & $\ket{l}$ & $\ket{l}$ & $ \ket{ll} $ & $\rho_{bb}$ \\
\hline
\end{tabular}
\label{tbl:CountProb}
\end{table*}
Following the scheme in figure~3 in the main text, we calculate the corresponding quantum dot density matrix $\rho_{\mathrm{dot}}(t)$ by solving the master equation with the fitted system parameters. Figure~\ref{fig:Dynamics} depicts exemplary dynamics of the quantum dot populations for different values of pump strength as a function of time, which is the result of solving the master equation. When the pulse hits the quantum dot, the biexciton level is excited and subsequently decays to the exciton level. There is also a detuned single photon transition from the ground state to the exciton level, which becomes more prominent for higher pump strength based on the broadening of spectral linewidth as a result of the intensity-dependent dephasing. For higher laser power, we observe an increased population of the dark state due to the laser dependent decay mechanisms to the dark states.

In order to estimate the photon density matrix $\rho_{\mathrm{photons}}$, we have to relate it to $\rho_{\mathrm{dot}}$. This is done by simulating the detector counts caused by the electric field emitted from the quantum dot as detailed below. Finally, $\rho_{\mathrm{photons}}$ can be constructed from the counts by using the tomographic method of the experiment~\cite{James:2001aa,Takesue:2009aa}.

\begin{figure}[t]
\centering
\includegraphics{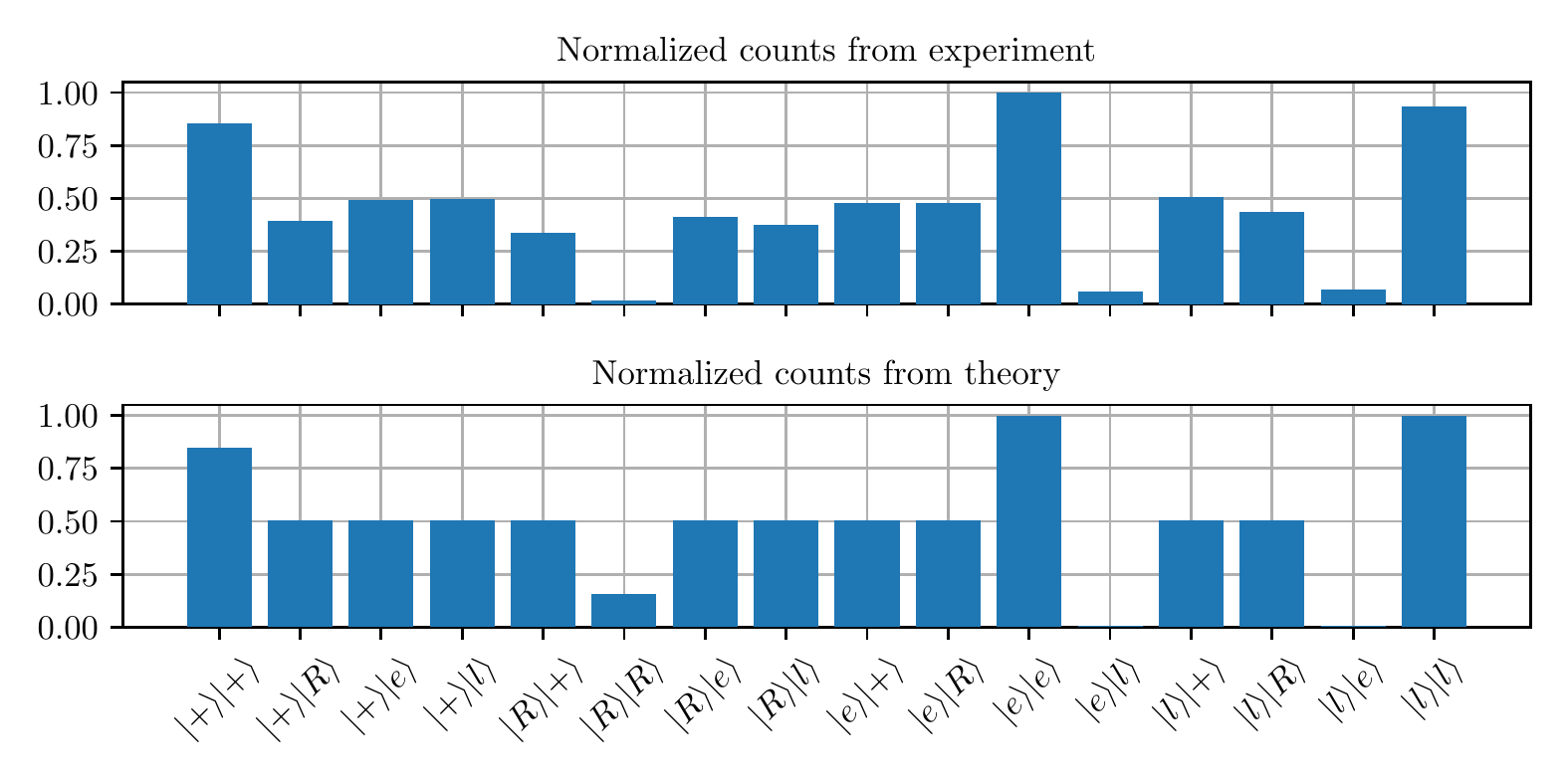}
\caption{\textit{Normalized counts for the 16 projective measurements obtained from experiment and theory.} The counts are normalized by the maximum of each data set. The horizontal axis marks the state for which we estimate the outcome of a projective measurement. To estimate the counts from theory, we use the expressions listed in table~\ref{tbl:CountProb}. By using the density matrix reconstruction theme, we derive the density matrices depicted in figure~4 in the main text from this data sets. }
\label{fig:ProjectiveCounts} 
\end{figure}
As shown in figure~1 in the main text, the detectors are faced with three photon peaks in time coming from the analysis interferometers. The first peak is the fraction of the early emission pulse traveling through the short interferometer arm, the third peak is due to the late pulse transmitted through the long arm and the second peak is a superposition of late and early emission. Detecting photons within the first and third peaks therefore corresponds to projective measurements on the $\ket{e}$ and $\ket{l}$ states. Detections in the second peak, in contrast, are projective measurements in a superposition basis, with their relative phases influenced by the analyzing interferometers. We consider the superposition states $\ket{+}=\frac{1}{\sqrt{2}} \Bigl( \ket{e} + \ket{l} \Bigr)$ and $\ket{R}=\frac{1}{\sqrt{2}} \Bigl( \ket{e} -i \ket{l} \Bigr)$.

In this way, we can realize projective measurements on the one-photon states $\{\ket{e},\ket{l},\ket{+},\ket{R}\}$. Combining those in the biexciton and exciton channel leads to a projection onto $16$ possible two-photon states (see table~\ref{tbl:CountProb} for an overview of those states). In order to be able to simulate these measurements, we relate an electric field to each peak at the output of the analysis interferometer. The expressions of the electric fields are given by the two dipole operators of the quantum dot. For the first peak, whose measurement corresponds to a projection on the early state $\ket{e}$, we get
\begin{align}
{}_{\ket{e}} E_{j}^{+}(t) \propto  \sigma_{j}^-(t').
\label{eq:FirstPeak}
\end{align}
As mentioned above, the second peak can be understood as a superposition of early and late emission. The expression of the electric field accounts for the phase of the superposition states $\ket{+}$ and $\ket{R}$,
\begin{align}
& {}_{\ket{+}} E_{j}^{+}(t) \propto  \frac{1}{\sqrt{2}} \Bigl( \sigma_{j}^-(t' + \Delta t) +  \sigma_{j}^- (t') \Bigr), \\
& {}_{\ket{R}} E_{j}^{+}(t) \propto  \frac{1}{\sqrt{2}} \Bigl( \sigma_{j}^-(t' + \Delta t) - i \sigma_{j}^- (t') \Bigr).
\end{align}
Finally, the expression for the last pulse reads
\begin{align}
& {}_{\ket{l}} E_{j}^{+}(t) \propto  \sigma_{j}^-(t' + \Delta t) \label{eq:ThirdPeak}.
\end{align}
We have introduced the notation such that $j \in \{ x,b \}$ with the dipole operators $\sigma_{x}^- = \ket{g} \bra{x}$ and $\sigma_{b}^- = \ket{x} \bra{b}$, $\Delta t$ being the time delay between the late and early emission and $t'=t-r/c$ with $r$ the distance between the quantum dot and the detector. A coincidence measurement between the biexciton and exciton channels corresponds to a projective measurement onto a two-photon state. The detection probability distribution of a coincidence measurement between detector $i$ and $j$, that is triggered at detector $i$, is given by~\cite{glauber_quantum_1963}
\begin{align}\label{eq:p}
&{}_{i,j}p_{\nu}(t) \propto  \langle {}_{1_{\nu}} \hat{E}_i^-  \ {}_{2_{\nu}} \hat{E}_j^- \ {}_{2_{\nu}} \hat{E}_j^+ \ {}_{1_{\nu}}  \hat{E}_i^+ \rangle .
\end{align}
Here, the index $\nu$ distinguishes the 16 two-photon states for which the expressions of the electric fields  have to be chosen according to the single-photon states that form the composite two-photon state. The indices $i,j \in \{ x,b \}$ distinguish the exciton and biexciton detection channel and $1_{\nu},2_{\nu} \in \{\ket{e}, \ket{l}, \ket{+}, \ket{R} \}$. An estimate of the coincidence counts $n_{\nu}$ can be calculated using
\begin{equation} \label{eq:n}
\begin{aligned}
n_{\nu} \propto	& \int {}_{x,b}p_{\nu}(t) + {}_{b,x}p_{\nu}(t) \ dt \\\
						& =: \int \eta_{\nu} (t) \ dt   .
\end{aligned}
\end{equation}
The time integral has to be taken over the time-span of the relevant dynamics. By substituting equations~\ref{eq:FirstPeak} to~\ref{eq:ThirdPeak} into equation~\ref{eq:p} we are left with calculating the expectation values of the dipole operators $\sigma^-$ and $\sigma^+$. In this way we can relate the coincidence counts $n_{\nu}$ to matrix elements of the quantum dot. See table~\ref{tbl:CountProb} for an overview of the two-photon states that are considered for the theoretical description and the corresponding expressions, which are proportional to $\eta_{\nu}$.

\begin{figure}[t]
\includegraphics{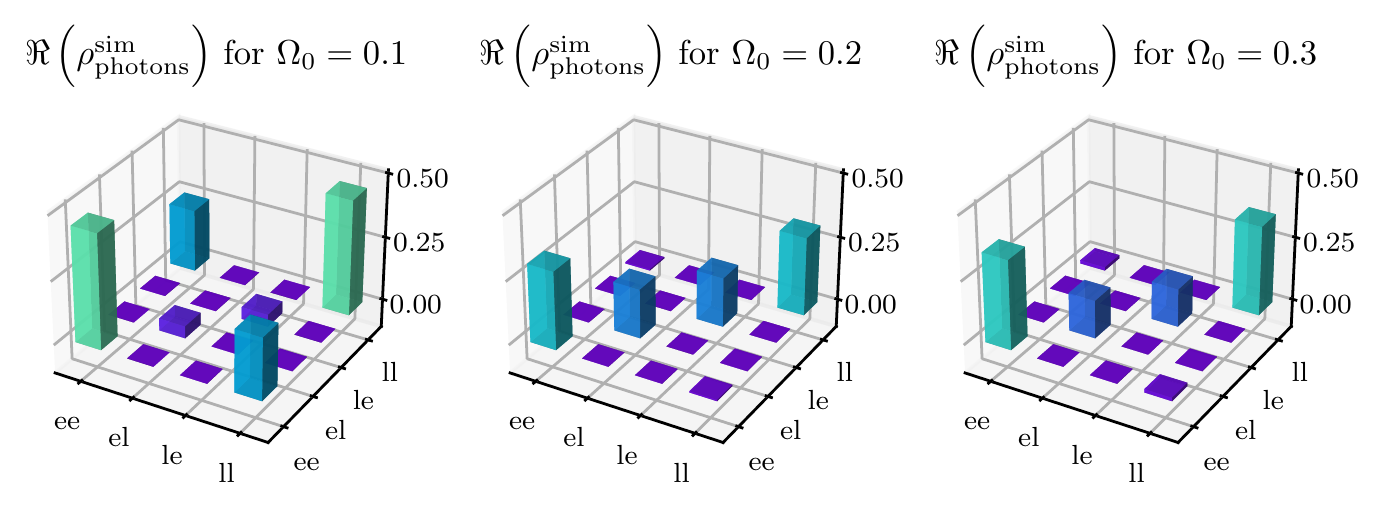}
\caption{\textit{Real part of simulated density matrix for various values of the Rabi frequency amplitude $\Omega_0$.} The laser pulse length for all three matrices is fixed at $85$ ps. We do not show the imaginary parts of these matrices as they are neglactably small.}
\label{fig:DensityMatrices_Omega}
\end{figure}
To reconstruct the photon density matrix $\rho_{\mathrm{photons}}$ from the number of counts $n_{\nu}$ we employ the experimental reconstruction method and use the relation~\cite{James:2001aa, Takesue:2009aa}
\begin{align} \label{eq:RhoPhotons}
\rho_{\mathrm{photons}} = \frac{1}{k} \sum_{\nu=1}^{16} M_{\nu} n_{\nu} \\
\text{with } k = \sum_{\nu=1}^{16} \text{tr}(M_{\nu}) n_{\nu}.
\end{align}
The transformation matrices $M_{\nu}$ depend on the chosen states on which the projective measurement is performed. They are defined by
\begin{align}
M_{\nu} := \sum_{i=1}^{16} \Gamma_i  B_{i,\nu}^{-1} .
\end{align}
The matrices $\Gamma_i$ can be seen as a basis for the two-photon matrices. They are given by
\begin{align}
\Gamma_i  := \sigma_{1_{i}} \otimes \sigma_{2_{i}} 
\end{align}
and represent any possible combination of the Pauli matrices and the identity matrix with
\begin{align}
\sigma_{j} \in \{ \mathds{1}, \sigma_{x}, \sigma_{y}, \sigma_{z}\}.
\end{align} 
The number $B_{i,\nu}^{-1}$ is the entry at row $i$ and column $\nu$ of the inverse matrix of $B$ whose elements in turn are defined via
\begin{align}
B_{i,\nu} := \bra{\psi_{i}} \Gamma_{\nu} \ket{\psi_{i}}.
\end{align}
Here, $\ket{\psi_{i}}$ are the two-photon states of the projective measurement.\\
To estimate the coincidence counts $n_\nu$ for the projective measurement on the two-photon state $\ket{\Psi_\nu}$ for all 16 states, we use the expressions proportional to $\eta_\nu$ given in table~\ref{tbl:CountProb} and use equation~\ref{eq:n}. We depict the estimated counts from theory and experiment for the 16 projective measurements in figure~\ref{fig:ProjectiveCounts}. The numerical estimates of $n_\nu$ can be fed into equation~\ref{eq:RhoPhotons} in order to reconstruct the density matrix of the photons $\rho_{\mathrm{photons}}$. Figure~4 in the main text shows the simulated density matrix for the system parameters gained from the set of parameter values obtained by the fit shown in figure~2b in the main text. Moreover, in figure~\ref{fig:DensityMatrices_Omega}, we depict the real part of additional density matrices for other values of average laser power but same pulse length compared to the chosen set of parameter values for the matrices in figure~4.

\begin{table}[ht]
\caption{\linespread{1.1}\selectfont Summary of the values used for the system parameters, which are fixed during the fitting loop for the Rabi oscillation data outlined in section~\ref{sec:RhoPhotonsEstimation}. The values are rounded to two decimal places.}
    \linespread{1.4}\selectfont
	\centering
	\begin{tabular}{m{0.4\textwidth}  m{0.1\textwidth}  m{0.32\textwidth}  }
		\hline
		Parameter & Symbol & Value \\ \hline \hline 
		Detuning between the exciton level and the laser & $\Delta_x$ & $1.60$ THz  \\ \hline
		Detuning between the biexciton level and the energy corresponding to the double laser frequency & $\Delta_b$ & $0$ THz (resonant excitation is assumed) \\ \hline 
		Measure for the width of the Rabi frequency (FWHM) & $\tau$ & $85$~ps  \\ \hline
		Biexciton decay rate & $\gamma_b$ & $1/458$~THz  \\ \hline
		Exciton decay rate & $\gamma_x$ & $1/1241$~THz  \\ \hline
		Exponent in expression of dephasing rate & $n$ & $2$ (``linear dephasing") \\ \hline
		Center of the laser pulse & $t_0$ & $300$~ps \\ \hline
		Total time for calculating the dynamics & $t_\mathrm{tot}$ & $7000$~ps \\ \hline
		Power offset & $k_{p_\mathrm{off}}$ & $0$ $\mu$W \\ \hline
	\end{tabular}
	\label{tbl:fixed_parameters}
\end{table}
\begin{table}[h]
\caption{\linespread{1.1}\selectfont Summary of parameters whose values are gained from the fit of the Rabi oscillations described in section~\ref{sec:RhoPhotonsEstimation}, given as rounded to two decimal places.}
    \linespread{1.4}\selectfont
	\centering
	\begin{tabular}{m{0.4\textwidth}  m{0.1\textwidth}  m{0.32\textwidth} }
		\hline
		Parameter & Symbol & Final value \\ \hline \hline 
		Biexciton-exciton laser dependent dephasing rate amplitude & $\gamma^{I_0}_{bx}$ & $0.03$~THz \\ \hline
		Exciton-ground state laser dependent dephasing rate amplitude & $\gamma^{I_0}_{xg}$ &$0.69$~THz \\ \hline
		Biexciton-exciton constant dephasing rate & $\gamma^{\text{const}}_{bx}$ & $0.56$~GHz \\ \hline		
		Exciton-ground state dephasing rate amplitude & $\gamma^{\text{const}}_{xg}$ & $0.25$~GHz \\ \hline
		Biexciton dark state decay rate amplitude & $\gamma^{I_0}_{bd}$ & $1.16$~GHz \\ \hline
		Exciton dark state decay rate amplitude & $\gamma^{I_0}_{xd}$ & $9.51$~GHz \\ \hline
		Power scaling factor & $k_{P_\mathrm{scale}}$ & $3.12\cdot 10^{6}$  \\ \hline
		Biexciton counts scaling factor & $k_{c_{\mathrm{scale}_b}}$ & $4.08 \cdot 10^{4}$  \\ \hline
		Exciton counts scaling factor & $k_{c_{\mathrm{scale}_x}}$ & $3.92 \cdot 10^{4}$  \\ \hline
		Biexciton counts offset  & $k_{c_{\mathrm{off}_b}}$ & $1.50\cdot 10^3$  \\ \hline
		Exciton counts offset  & $k_{c_{\mathrm{off}_x}}$ & $1.50\cdot 10^3$  \\ \hline
	\end{tabular}
	\label{tbl:fit_parameters}
\end{table}
\section{System Parameter values}\label{sec:SystemParameters}

\begin{figure}[t]
\includegraphics{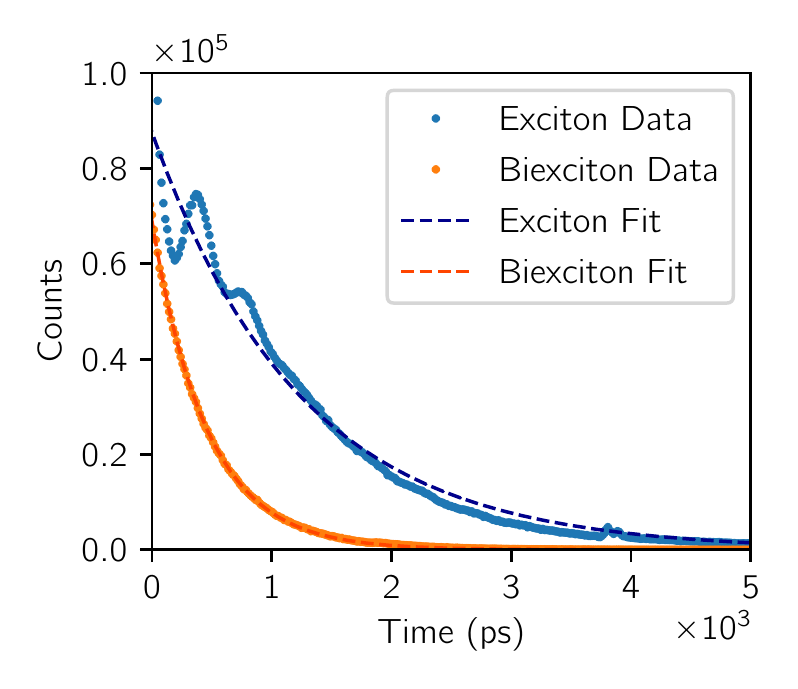}
\caption{\textit{Decay data and fit of the biexciton and exciton level.}
The decaying counts of the biexciton and exciton photons are fitted to an exponential decay. From the exponent of the fitted function, we retreive the decay rates of the biexciton and exciton level without considering the losses to the dark state.}
\label{fig:Decay}
\end{figure}

To compare the measurements in the experiment to our simulation of the system and to calibrate the theoretical model, we use five different sets of data. First, we use the data for the exponential decay of the biexciton and exciton, which is depicted in figure~\ref{fig:Decay}, to obtain the corresponding decay rates $\gamma_b$ and $\gamma_x$ from the exponents of the fit. We do not consider the additional intensity dependent loss to the dark state for this procedure. The resulting values are given in table~\ref{tbl:fixed_parameters}. The second set of data consists of the quantum dot's emission spectrum (see figure~\ref{fig:spectrum}) from which we retrieve the exciton detuning (see table~\ref{tbl:fit_parameters}). The similar brightness of the sharp emission lines in the spectrum is a result of directly populating the biexciton state via a two-photon resonant excitation process and a resulting photon emission cascade into the ground state. Third, the signal from the autocorrelator leads to the pulse length given in table~\ref{tbl:fit_parameters}. The fourth set of data is constituted by the Rabi oscillations depicted in figure~2 in the main text. We obtain the values listed in table~\ref{tbl:fit_parameters} by calibrating our model with the Rabi oscillations within a comprehensive fitting loop which is outlined in section~\ref{sec:RhoPhotonsEstimation}. Finally, the fifth set of parameters is represented by the number of counts for the 16 projective measurements, which are covered in section~\ref{sec:RhoPhotonsEstimation} (see figure~\ref{fig:ProjectiveCounts}). We use these counts to reconstruct the density matrix of the photons to compare the matrix predicted by the model to the one gained from experimental data.

\section{Derivation of the Hamiltonian}\label{sec:Hamiltonian}
In this section we will demonstrate how the Hamiltonian, which is introduced in the main text, can be derived from a more basic description of the system. This derivation can be found in similar form in reference~\cite{aumann_tomography_2017} as well. We start with a Hamiltonian in the dipole representation for a three-level system that is driven by a monochromatic light source:

\begin{align} \label{eq:HamiltonianDipoleRepresentation}
H = E_g \ket{g}\bra{g} + E_x \ket{x}\bra{x} + E_b \ket{b}\bra{b} - \vec{d}\vec{E}_{\text{cl}} (t) := H_{0} + H_{\textrm{int}},
\end{align}
with the dipole operator $\vec{d}$. The ground state energy can be set to $0$ and the remaining energy scales can be expressed by the laser frequency $\omega_L$ and the detunings $\Delta_x$ and $\Delta_b$, such that $E_x=\omega_L + \Delta_x$ and $E_b = 2\omega_L - \Delta_b $. We note, that we set $\hbar =1 $ for the analytical calculations in this article. For the electric laser field we use the expression:
\begin{align}
    \vec{E}_{\text{cl}} = \vec{E}_{L} (t) \left( e^{i \omega_L t} + e^{-i \omega_L t} \right).
\end{align}
We introduce the notation $\bra{i} \vec{d} \ket{j} =: \vec{d}_{ij}$ and $\bra{i} \Omega \ket{j} := \Omega_{ij}$ with the Rabi frequency $\Omega$. The Rabi frequency is associated with the electric field by
\begin{align}
    \Omega_{ij}(t) = -\vec{d}_{ij} \vec{E}_{L} (t).
\end{align}
We assume the same coupling between the ground state and exciton state and between the exciton state and biexciton state, which leads to 
\begin{align}
    \Omega := \Omega_{gx} = \Omega_{xb}.
\end{align}
The Rabi frequencies are assumed to be real (i.e. $\Omega_{ij} = \Omega_{ij}^* =\Omega_{ji} \ \forall i,j \in {x,g,b}$) and we further assume no direct coupling between the ground state and biexciton state:
\begin{align}
    \Omega_{gb} = \Omega_{bg} = 0.
\end{align}
For a reduced expression of the interaction Hamiltonian, we require the diagonal elements of the dipole operator to vanish. In this case we have:
\begin{align}
\hat{H}_{int} = - \vec{d} \vec{E}_{\text{cl}} (t) = \Omega (t) \ (e^{i \omega_{\text{L}} t} + e^{- i \omega_{\text{L}} t}) (\ket{g}\bra{x} + \ket{x}\bra{b} + \text{h.c.}). 
\end{align}
We can transform the Hamiltonian, such that factors appear which oscillate with double the laser frequency. The chosen unitary transformation is  $U := e^{i (\omega_{\text{L}}+\Delta_b) t \ket{x}\bra{x} } e^{i (2 \omega_{\text{L}} + \Delta _b ) t \ket{b}\bra{b} }$. With that trasformation we get:
\begin{align}
\widetilde{H} &:= U H U^{\dagger} + i (\partial _t U)  U^{\dagger} \\
&= (\Delta _x - \Delta_b) \ket{x}\bra{x} - 2 \Delta_b \ket{b}\bra{b} +  \Omega (t) \left( \left(1+e^{-i2\omega_{L} t}\right) \left( e^{-i \Delta_b t} \ket{g}\bra{x} +  \ket{x}\bra{b}\right) + \text{h.c.}\right).
\end{align}
To further reduce the expression, we drop the terms that oscillate with double the laser frequency $e^{-i2\omega_{L} t}$ and it's complex conjugate. The factor $e^{-i \Delta_b t}$ can be absorbed in the state $\ket{g}$, i.e. $e^{-i \Delta_b t}\ket{g} \rightarrow{\ket{g}}$. With these transformations, we obtain the Hamiltonian which is given in the main text:
\begin{align}
\widetilde{H} = (\Delta _x - \Delta_b) \ket{x}\bra{x} - 2 \Delta_b \ket{b}\bra{b} +  \Omega (t) \left(\ket{g}\bra{x} +  \ket{x}\bra{b} + \text{h.c.}\right).
\end{align}

\FloatBarrier
\section*{References}
\bibliographystyle{unsrt}